\documentclass[fleqn,usenatbib]{mnras}

\usepackage{newtxtext,newtxmath}

\usepackage[T1]{fontenc}

\DeclareRobustCommand{\VAN}[3]{#2}
\let\VANthebibliography\thebibliography
\def\thebibliography{\DeclareRobustCommand{\VAN}[3]{##3}\VANthebibliography}

\usepackage{graphicx}	
\usepackage{amsmath}	

\usepackage{amssymb}	
\usepackage{longtable}

\defcitealias{2023ApJS..266...28L}{Paper I}
\defcitealias{2024ApJ...962...44L}{Paper II}

\title[Pulsation Phases and Mode in Kepler Heartbeat Stars]{Pulsation Phases and Mode Identification of Tidally Excited Oscillations in Fourteen Kepler Heartbeat Stars}

\author[Li et al.]{
Min-Yu Li,$^{1,2,3,4}$
Sheng-Bang Qian,$^{2,3}$\thanks{E-mail: qsb@ynao.ac.cn}
Li-Ying Zhu,$^{1,4}$\thanks{E-mail: zhuly@ynao.ac.cn}
Zhao Guo, $^{5}$
Wen-Ping Liao,$^{1,4}$
\newauthor
Er-Gang Zhao,$^{1,4}$
Xiang-Dong Shi,$^{1,4}$
Fu-Xing Li,$^{1,4}$
Qi-Bin Sun,$^{1,4}$
\\
$^{1}$Yunnan Observatories, Chinese Academy of Sciences, Kunming 650216, People's Republic of China\\
$^{2}$Department of Astronomy, School of Physics and Astronomy, Yunnan University, Kunming 650091, People's Republic of China\\
$^{3}$Key Laboratory of Astroparticle Physics of Yunnan Province, Yunnan University, Kunming 650091, People's Republic of China\\
$^{4}$University of Chinese Academy of Sciences, No.1 Yanqihu East Road, Huairou District, Beijing 101408, People's Republic of China\\
$^{5}$Department of Applied Mathematics and Theoretical Physics, University of Cambridge, Cambridge CB3 0WA, UK
}

\date{Accepted XXX. Received YYY; in original form ZZZ}

\pubyear{2023}

\begin{document}
\label{firstpage}
\pagerange{\pageref{firstpage}--\pageref{lastpage}}
\maketitle

\begin{abstract}
Tidally excited oscillations (TEOs) in Heartbeat Stars (HBSs) are an essential probe of the internal properties of the systems, but their potential has yet to be fully exploited. Based on the orbital parameters of TEO candidates from our previous works, we identify the pulsation phases and amplitudes of TEOs in fourteen Kepler HBSs. Most pulsation phases of most systems can be explained by the dominant being $l=2$, $m=0$, or $\pm2$ spherical harmonic, assuming that the spin and orbital axes are aligned, and the pulsations are adiabatic and standing waves. The largest deviation ($>6\sigma$) occurs in KIC 8459354, which can be explained by the spin-orbit misalignment, and KIC 5877364 has a similar scenario. For KIC 11122789, almost half of the harmonics show large deviations; we cautiously suggest that these harmonics may not be considered TEO candidates. A similar scenario also exists in KIC 6290740. This phases and mode identification approach can also be used inversely to verify the TEO candidates derived by the Fourier analysis. Furthermore, the harmonics with large deviations ($>2\sigma$) in KIC 4377638, KIC 5090937, and KIC 11403032 can be expected to be travelling waves rather than standing waves. In addition, we also suggest that the apsidal motion could cause large deviations in TEO phases from theoretical values.

\end{abstract}




\section{Introduction}

Tidally excited oscillations (TEOs) are induced by the phase-dependent tides in binary stars with eccentric orbits \citep{1975A&A....41..329Z,1995ApJ...449..294K,2017MNRAS.472.1538F,2021FrASS...7..102J,2022A&A...659A..47K}. They can occur in some of the Heartbeat Stars (HBSs) and in turn affect the evolution of the HBSs, including the synchronization and circularization of binary stars. Therefore, TEOs can be a probe to the binary system parameters and potentially to the stellar interiors \citep{2020ApJ...888...95G}. Theoretical work was done several decades ago \citep{1975A&A....41..329Z,1995ApJ...449..294K}, but it is the Kepler satellite \citep{2010Sci...327..977B} that provides the opportunity to study such small variations in photometric light curves. A series of works have reported some studies of TEOs in the Kepler HBSs.

The Kepler HBS KOI-54 (KIC 8112039) was first reported by \citet{2011ApJS..197....4W}, with the two dominant TEOs being the 90th and 91st harmonics of the orbital frequency. KOI-54 was then studied in a series of papers. \citet{2012MNRAS.420.3126F} studied the nature of the 90th and 91st harmonics and suggested that the system could naturally evolve to a state where at least one $m=2$ mode is locked in resonance with the orbit. However, \citet{2012MNRAS.421..983B} found that chance resonance of $l=2$, $m=0$ modes can also explain the observed large amplitude of the 90th and 91st harmonics without invoking resonance locking. \citet{2014MNRAS.440.3036O} also found that the oscillation modes are axisymmetric and thus cannot be responsible for resonance locking. Most of the high frequency harmonic pulsations are found to be $m=0$ oscillations. Recently, \citet{2022MNRAS.517..437G} reexamined the non-linear TEOs in KOI-54 and found that the non-orbital-harmonic TEOs show a period spacing pattern, likely arising from $l=2$, $m=0$ eigen-modes. The two largest-amplitude TEOs at 90th and 91st harmonics are very close to resonances, and likely from different stars. \citet{2017ApJ...834...59G, 2019ApJ...885...46G} modeled both the amplitudes and phases of TEOs in the eclipsing HBSs KIC 3230227 and KIC 4142768, respectively. They found that most TEO phases in the two systems are consistent with $l=2$, $|m|=2$ modes.

\citet{2020ApJ...888...95G} studied the pulsation phases of TEOs in eight HBSs and found that more than half of the systems satisfy the geometric effect of the dominant $l=2$, $m=0$, or $\pm2$ modes assuming their spin and orbital axes are aligned and the pulsations are adiabatic. The phases of TEOs contain important information about the mode properties and can be used to better understand the mode physics of HBSs. However, the detailed analysis of pulsation phases and mode identification in more HBSs with TEOs is lacking due to the need for better characterization of the orbital and stellar parameters. 

In the first paper in the series (\citet{2023ApJS..266...28L}, hereafter \citetalias{2023ApJS..266...28L}), we derived the orbital parameters of 153 Kepler HBSs in the \citet{2016AJ....151...68K} catalog by fitting a corrected version of the \citet{1995ApJ...449..294K} model (K95$^+$ model). We then identified the TEOs in 21 HBSs among these systems in the second paper in the series (\citet{2024ApJ...962...44L}, hereafter \citetalias{2024ApJ...962...44L}). So far, we have obtained sufficient orbital and stellar parameters of the HBSs with TEOs based on our previous works. Therefore, we can study the pulsation phases and mode identification of these TEOs.

This paper examines the pulsation phases of TEOs from fourteen HBSs following the analysis approach of \citet{2014MNRAS.440.3036O} and \citet{2020ApJ...888...95G}. Section \ref{sec:analysis} describes the equations and the analytic procedure. Section \ref{sec:rst} presents the results. Section \ref{sec:discussion} discusses the deviations of the harmonics and addresses another potential cause of the deviations. Section \ref{sec:conclusions} summarizes and concludes our work.

\section{Analysis} \label{sec:analysis}
Under the following assumptions: (1) the pulsation, spin, and orbit axes are all aligned; (2) the pulsations are adiabatic and the observed TEOs are standing waves; (3) the observed TEOs are not fine-tuned, the pulsation phases of the TEOs for dominant modes of spherical harmonic degree $l=2$ can be expressed by the equation \citep{2014MNRAS.440.3036O, 2020ApJ...888...95G}:
\begin{equation} \label{equation:phi}
	\phi_{_{l=2,m}}=0.25+m\phi_{_{0}},
\end{equation}
where azimuthal order $m = 0, \pm2$, $\phi_{_{0}}=0.25-\omega/360^{\circ}$ and $\omega$ is the argument of periastron. In addition, all phases can be measured with respect to the time of periastron passage $T_{_{0}}$ and are in units of 360$^{\circ}$.

According to Eq. (\ref{equation:phi}), the accurate $\omega$ and $T_{_{0}}$ are needed to identify the pulsation modes from their phases. Since the parameters of these TEO HBSs have been derived in \citetalias{2023ApJS..266...28L}, the theoretical phases of the $m = 0, \pm2$ modes of these harmonic TEOs can be calculated. The phases derived from the Fourier spectrum can then be used to identify the azimuthal order $m$. 

However, note that the $\omega$ derived from different approaches may differ in some systems. We will perform a comprehensive analysis of these differences. In the cases, including KIC 3547874, KIC 5090937, KIC 5960989, and KIC 11403032, we further extend the range of theoretical phases $\phi$ for $m=\pm2$ modes to account for the different $\omega$ derived by different approaches.

In addition, \citet{2020ApJ...888...95G} have investigated the amplitude ratio as a function of orbital inclination $i$ (see their Figure 1). They have concluded that a low inclination (less than 30$^{\circ}$) would favor the $m = 0$ mode, while medium to high inclinations would favor both the $m = 0$ and $m = 2$ modes. These rules can also further constrain the identified modes.

To derive the phases of the harmonic frequencies, we perform a standard prewhitening procedure using Period04 \citep{2005CoAst.146...53L}, a software package based on classical Fourier analysis. The flux variation is formulated as a sinusoidal function $A_i {\rm sin} [2\pi(f_i \cdot t+\phi_i)]$, where $A_i$, $f_i$, and $\phi_i$ are the amplitude, frequency, and phase, respectively. The uncertainties of the three parameters are obtained using Period04, which calculates the uncertainties according to \citet{1999DSSN...13...28M}. In addition, since $t$ is measured relative to the time of periastron $T_{_{0}}$, the time of each data point in the photometric light curves should subtract $T_{_{0}}$ before Fourier analysis. The mean noise level of each frequency, $N_i$, is derived as the mean amplitude in the frequency range $\pm$0.5 d$^{-1}$. The signal-to-noise ratio ($S/N$) of the frequencies is $A_i / N_i$. 

Consider KIC 3547874 as an example, shown in Figure \ref{fig:3547874}. The top panel shows the K95$^+$ model fitted to the phase-folded light curve, and all the orbital parameters are derived in \citetalias{2023ApJS..266...28L}. The middle panel shows the Fourier spectrum of the TEOs analysis results derived in \citetalias{2024ApJ...962...44L}. The harmonic numbers $n$ of the TEOs are plotted as blue dashed lines. The bottom panel shows the pulsation phases of the TEOs. The theoretical pulsation phases $\phi_{_{m}}$ for $m = 0, \pm2$ strips are derived by Eq. (\ref{equation:phi}). In addition, since the phases alone cannot determine whether the modes are retrograde ($m=+2$) or prograde ($m=-2$) due to the 180$^{\circ}$ phase ambiguity \citep{2012MNRAS.421..983B}, a phase offset of 0.5 (180$^{\circ}$) can be added or subtracted from Eq. (\ref{equation:phi}) when comparing with observations \citep{2020ApJ...888...95G}. Therefore, we obtain two strips for each $m = 0, \pm2$ modes. The width of the strips result from the uncertainties of $T_{_{0}}$ and $\omega$ \citep{2014MNRAS.440.3036O} that are obtained from \citetalias{2023ApJS..266...28L}. However, for this system, the $\omega$=40$^\circ$.68 derived by \citet{2017MNRAS.469.2089D} has a more significant difference from ours. We then plot the dashed lines to represent the theoretical phase calculated by the more significant difference $\omega$. The lighter-colored areas show the total uncertainties. Each TEO is represented by a blue circle. Its position corresponds to its phase and harmonic number $n$; the size of each circle corresponds to the amplitude of the frequency; the error bar corresponds to the uncertainty of the phase. In Table \ref{tab:3547874}, the orbital parameters are from \citetalias{2023ApJS..266...28L}, and the frequency, amplitude, and phase of each harmonic are derived in this work.

Among the HBSs reported in \citetalias{2024ApJ...962...44L}, we exclude the systems with $\omega$ close to 90$^\circ$, since the $m = 0, \pm2$ strips will be too close to distinguish. We also exclude the systems with lower $S/N$ TEOs, since the error bars of the phases will be too large to match the modes. Finally, we select the fourteen samples discussed in Section \ref{sec:rst}.

\section{Results}\label{sec:rst}
We divide the results into three groups. Section \ref{sec:meet} presents seven systems where all pulsation phases meet or close to the adiabatic expectations. In Section \ref{sec:misaligned}, two systems are shown to have a misaligned spin orbit in which most of the pulsation phases have large deviations from the adiabatic expectations. Section \ref{sec:part_meet} presents five systems whose pulsation phases are partially consistent with adiabatic expectations. Among them, the harmonics with large deviations in the first two systems cannot be TEO candidates, and the deviation modes in the other three systems can be expected to be travelling waves rather than standing waves.

\subsection{Systems that meet or close to expectations}\label{sec:meet}
\begin{figure}
	\includegraphics[width=\columnwidth]{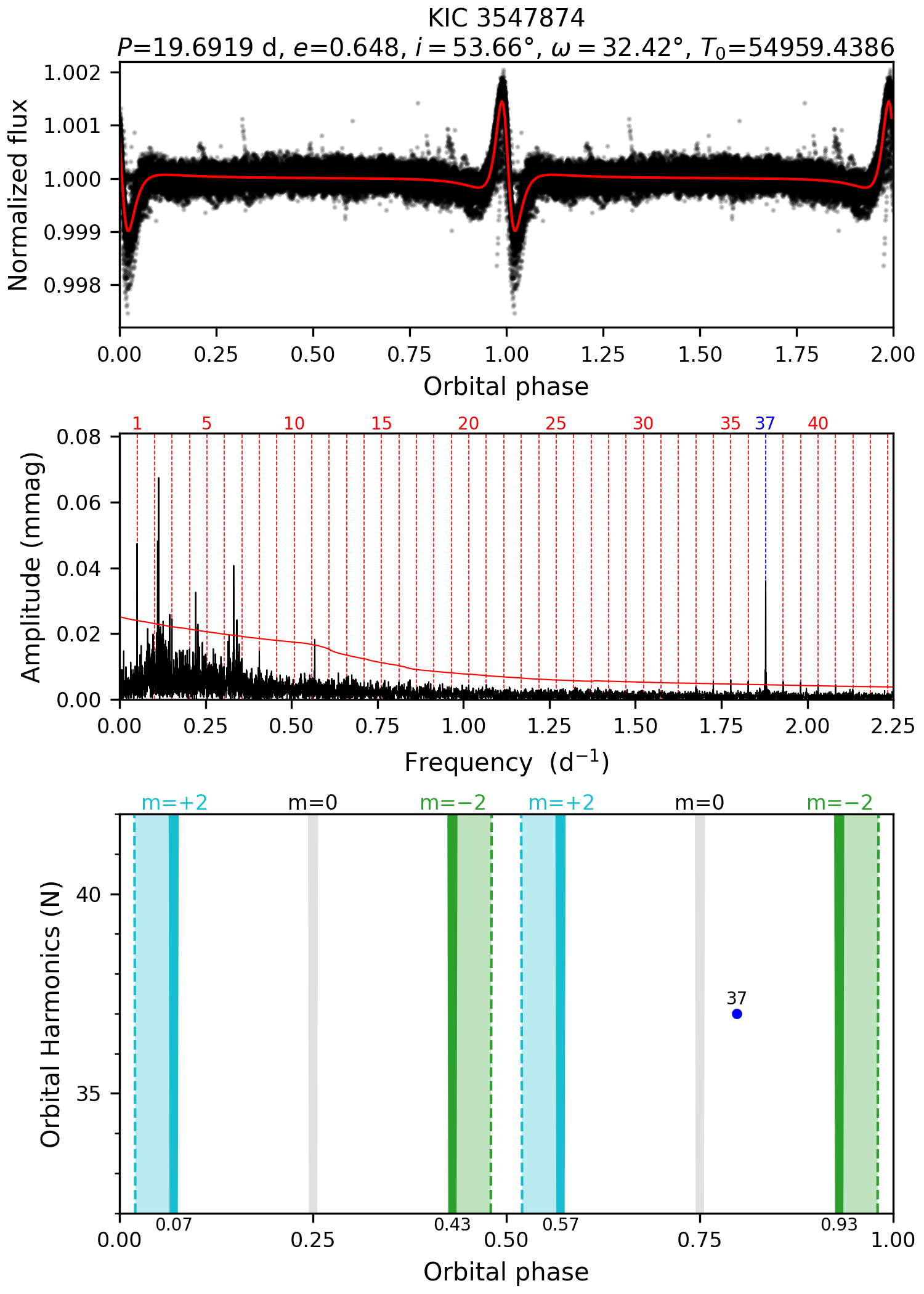}
	\caption{Top panel: The K95$^+$ model (solid red line) fitted to the phase-folded light curve (black dots). Middle panel: The Fourier spectrum of the TEOs analysis results. The red and blue vertical dashed lines represent the orbital harmonics n; the blue ones indicate that they are TEOs. The solid red line represents the amplitudes at $S/N$ = 4.0 as a function of frequency. Bottom panel: The pulsation phases of the TEOs. The gray, light blue, and green strips indicate the $m=0,+2$, and $-2$ modes, respectively. The phases of the $m=+2,-2$ modes are shown under the center of the strips. The dashed lines represent the theoretical phase calculated from the $\omega$ with a larger difference; the lighter-colored areas indicate the total uncertainties. The blue circle represents a TEO with its harmonic number $n$; the size corresponds to its amplitude; the error bar corresponds to the uncertainty of its phase.
		\label{fig:3547874}}
\end{figure}

\subsubsection{KIC 3547874 (Figure \ref{fig:3547874}; Table \ref{tab:3547874})}
\citet{2012ApJ...753...86T} reported that KIC 3547874 is a double-lined HBS with $e$=0.648, $i$=46$^\circ$.4, and $\omega$=40$^\circ$.40 based on the K95$^+$ model. They also reported that the TEOs are present in this HBS. \citet{2017MNRAS.469.2089D} obtained $e$=0.655, a higher inclination $i$=67$^\circ$.7, and a similar argument of periastron $\omega$=40$^\circ$.68 using PHOEBE. We obtained an eccentricity of 0.648 and an inclination of 53$^{\circ}$.7 and $\omega$=32$^\circ$.424 from \citetalias{2023ApJS..266...28L}. Note that the $\omega$ has a difference of about 8$^\circ$ from that of \citet{2012ApJ...753...86T} and \citet{2017MNRAS.469.2089D}. The dashed lines in the bottom panel represent the theoretical pulsation phases of $\omega$=40$^\circ$.68, and the lighter-colored strips represent the corresponding differences. The $n$ = 37 prominent TEO stands out clearly in the middle panel of Figure \ref{fig:3547874}. The bottom panel shows that its pulsation phase is close to an $m = 0$ mode. In addition, the heartbeat signal near the periastron are difficult to model well (see Figure A1 in \citetalias{2024ApJ...962...44L}); this may cause the differences in $\omega$ between different works. However, the $m = 0$ mode can favor such an intermediate inclination system.

\begin{table}
	\caption{Orbital and TEO parameters of KIC 3547874.}
	\label{tab:3547874}
	\begin{tabular*}{\columnwidth}{l@{\hspace*{9pt}}l@{\hspace*{9pt}}l@{\hspace*{8pt}}l@{\hspace*{7pt}}l}
		\hline
		Orbital Parameter &  & \\
		\hline
		$P$(d) & 19.6919178(70) & \\
		$e$ & 0.64755(10) & \\
		$i$($^\circ$) & 53.663(18) & \\
		$\omega$($^\circ$) & 32.424(42) & \\
		$T_{_{0}}$(BJD$-$2,400,000) & 54959.43861(47) & \\
		\hline
		Harmonic number & Frequency & Amplitude & Phase & S/N \\
		n & (d$^{-1}$) & (mmag)  &  \\
		37 & 1.878924(10) & 0.03600(98) & 0.798(4) & 32.44 \\ 
		\hline
	\end{tabular*}
\end{table}

\begin{figure}
	\includegraphics[width=\columnwidth]{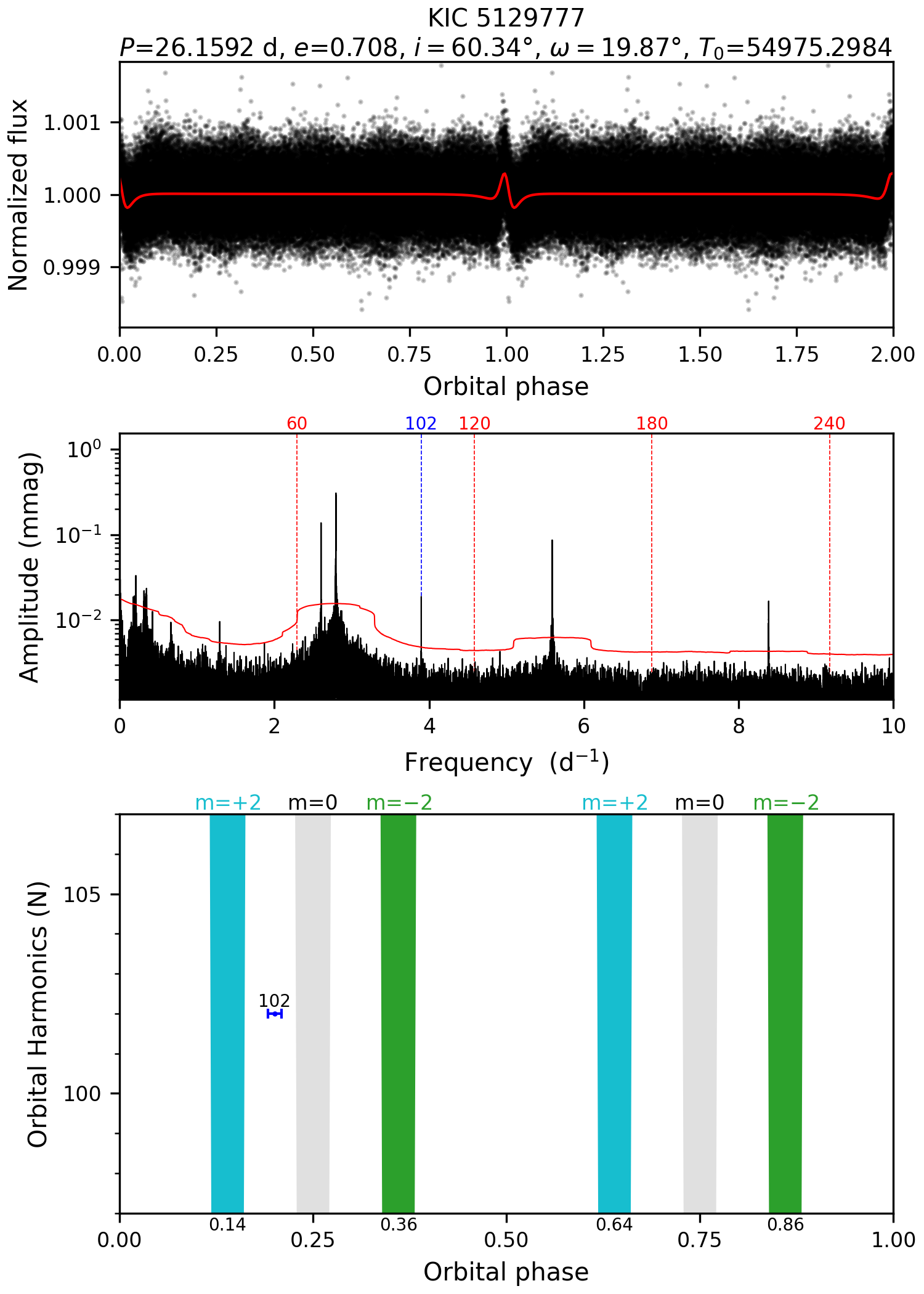}
	\caption{Same as Fig. \ref{fig:3547874} for KIC 5129777.
		\label{fig:5129777}}
\end{figure}

\subsubsection{KIC 5129777 (Figure \ref{fig:5129777}; Table \ref{tab:5129777})}
This is a 26 day binary system, with an eccentricity of 0.71, an inclination of 60$^{\circ}$, and an argument of periastron $\omega$ of 20$^\circ$, as shown in Figure \ref{fig:5129777}. The $n$ = 102 harmonic is clearly visible in the middle panel. The bottom panel shows that its pulsation phase is close to the $m = 0$ mode. However, since the heartbeat signal has a small amplitude in the light curves, the uncertainty of $\omega$ may be underestimated by the K95$^+$ model, resulting in its pulsation phase also being close to the $m=2$ mode. Other modeling approaches may derive a more accurate $\omega$ in future work.

\begin{table}
	\caption{Orbital and TEO parameters of KIC 5129777.}
	\label{tab:5129777}
	\begin{tabular*}{\columnwidth}{l@{\hspace*{9pt}}l@{\hspace*{9pt}}l@{\hspace*{8pt}}l@{\hspace*{7pt}}l}
		\hline
		Orbital Parameter &  & \\
		\hline
		$P$(d) & 26.15918(18) & \\
		$e$ & 0.7080(15) & \\
		$i$($^\circ$) & 60.34(46) & \\
		$\omega$($^\circ$) & 19.87(72) & \\
		$T_{_{0}}$(BJD$-$2,400,000) & 54975.2984(86) & \\
		\hline
		Harmonic number & Frequency & Amplitude & Phase & S/N \\
		n & (d$^{-1}$) & (mmag)  &  \\
		102 & 3.899293(21) & 0.0188(10) & 0.200(19) & 15.53 \\ 
		\hline
	\end{tabular*}
\end{table}

\begin{figure}
	\includegraphics[width=\columnwidth]{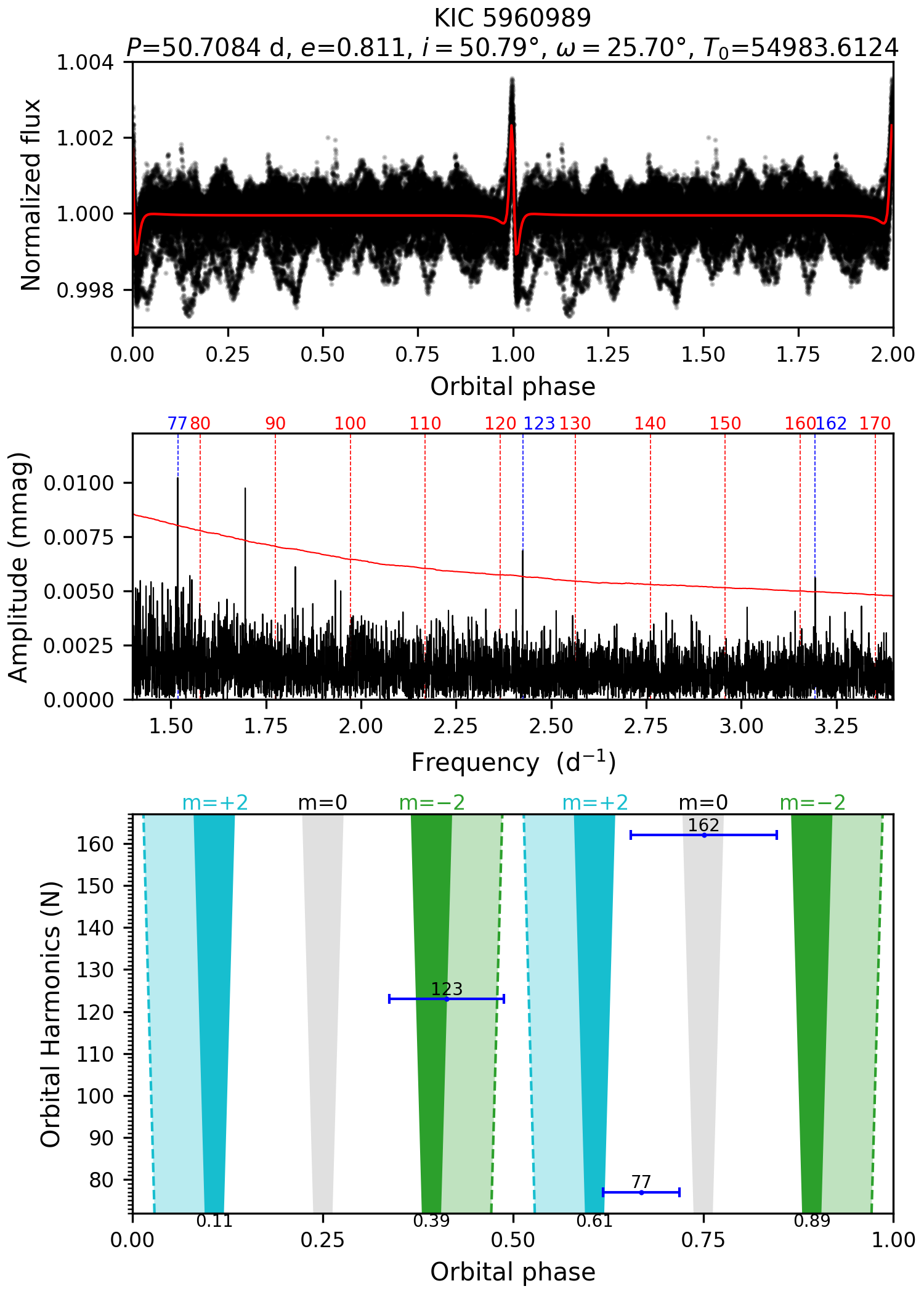}
	\caption{Same as Fig. \ref{fig:3547874} for KIC 5960989.
		\label{fig:5960989}}
\end{figure}

\subsubsection{KIC 5960989 (Figure \ref{fig:5960989}; Table \ref{tab:5960989})}
\citet{2016ApJ...829...34S} reported a long orbital period of 50 days, a high eccentricity $e$=0.813, and an argument of periastron $\omega$ of 37$^\circ$.873. \citet{2017MNRAS.469.2089D} obtained $e$=0.806, an inclination $i$=51$^\circ$, and an argument of periastron $\omega$ of 29$^\circ$.82.  We obtained an eccentricity of 0.811, an inclination of 50$^{\circ}$.8, and $\omega$=25$^\circ$.7 from \citetalias{2023ApJS..266...28L}. The bottom panel shows that the $n$ = 77, 123, and 162 harmonics are close to the $m=2, -2$, and $0$ modes, respectively. The dashed lines represent the theoretical pulsation phases of $\omega$=37$^\circ$.873, and the lighter-colored strips represent the corresponding deviations.

\begin{table}
	\caption{Orbital and TEO parameters of KIC 5960989.}
	\label{tab:5960989}
	\begin{tabular*}{\columnwidth}{l@{\hspace*{9pt}}l@{\hspace*{9pt}}l@{\hspace*{8pt}}l@{\hspace*{7pt}}l}
		\hline
		Orbital Parameter &  & \\
		\hline
		$P$(d) & 50.708376(41) & \\
		$e$ & 0.81126(15) & \\
		$i$($^\circ$) & 50.792(40) & \\
		$\omega$($^\circ$) & 25.70(11) & \\
		$T_{_{0}}$(BJD$-$2,400,000) & 54983.6124(10) & \\
		\hline
		Harmonic number & Frequency & Amplitude & Phase & S/N \\
		n & (d$^{-1}$) & (mmag)  &  \\
		77 & 1.51838(12) & 0.0102(32) & 0.669(50) & 5.02 \\ 
		123 & 2.42553(18) & 0.0068(32) & 0.413(75) & 4.74 \\ 
		162 & 3.19463(23) & 0.0054(32) & 0.751(96) & 4.30 \\ 
		\hline
	\end{tabular*}
\end{table}

\begin{figure}
	\includegraphics[width=\columnwidth]{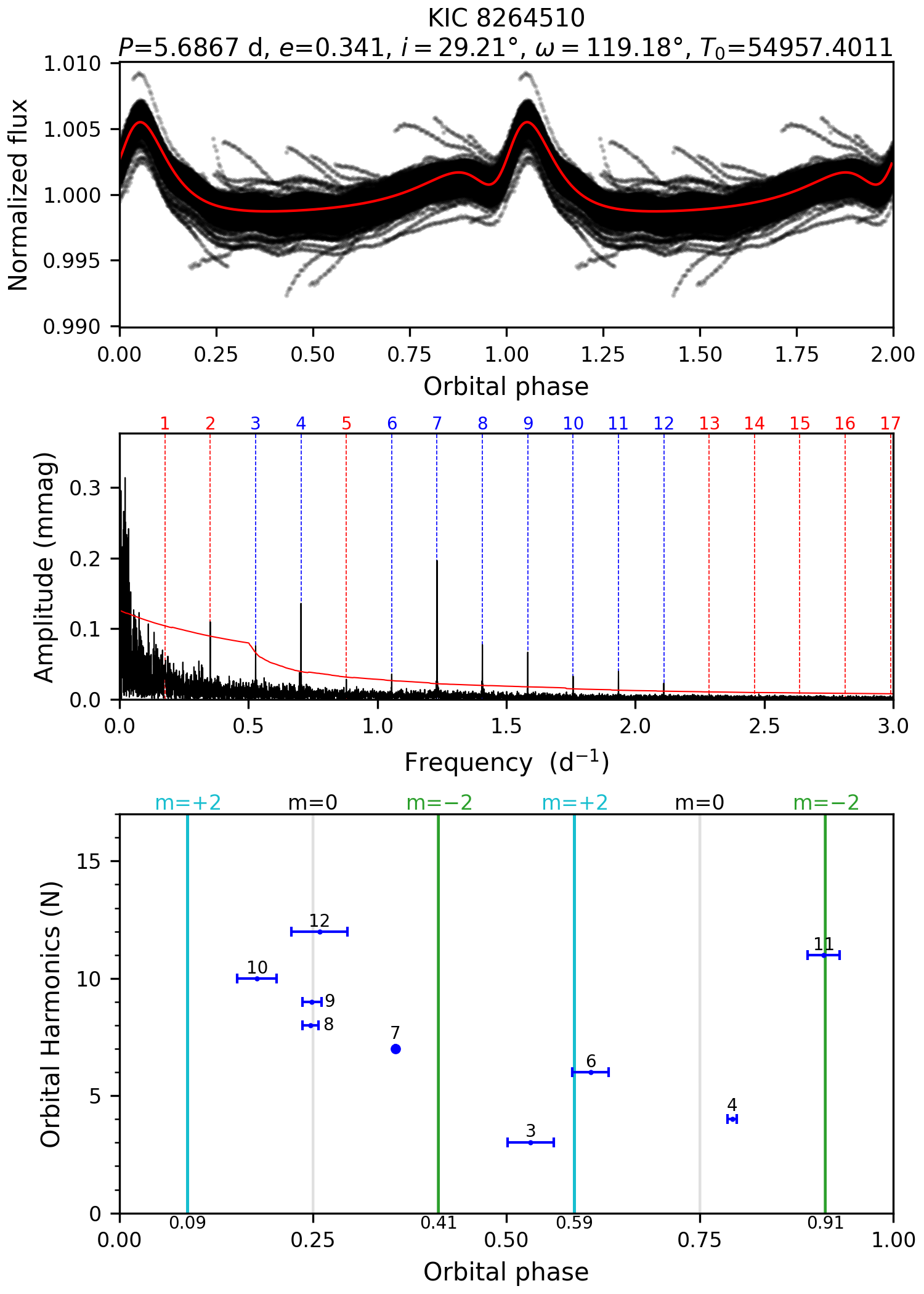}
	\caption{Same as Fig. \ref{fig:3547874} for KIC 8264510.
		\label{fig:8264510}}
\end{figure}

\subsubsection{KIC 8264510 (Figure \ref{fig:8264510}; Table \ref{tab:8264510})}
\citet{2012ApJ...753...86T} reported the eccentricity $e$=0.338, the inclination $i$=29$^\circ$, and the argument of periastron $\omega$ of 120$^\circ$. They also reported the presence of TEOs in this HBS. We obtained an eccentricity of 0.34, an inclination of 29$^{\circ}$.2, and an argument of periastron $\omega$ of 119$^\circ$.2 from \citetalias{2023ApJS..266...28L}. The $n$=4, 8, 9, 10, and 12 harmonics are consistent with or close to the $m=0$ mode; the $n$=3, 6, 7, and 11 harmonics are close to the $|m|=2$ modes.

\begin{table}
	\caption{Orbital and TEO parameters of KIC 8264510.}
	\label{tab:8264510}
	\begin{tabular*}{\columnwidth}{l@{\hspace*{9pt}}l@{\hspace*{9pt}}l@{\hspace*{8pt}}l@{\hspace*{7pt}}l}
		\hline
		Orbital Parameter &  & \\
		\hline
		$P$(d) & 5.68674956(31) & \\
		$e$ & 0.34111(3) & \\
		$i$($^\circ$) & 29.211(1) & \\
		$\omega$($^\circ$) & 119.179(4) & \\
		$T_{_{0}}$(BJD$-$2,400,000) & 54957.401056(55) & \\
		\hline
		Harmonic number & Frequency & Amplitude & Phase & S/N \\
		n & (d$^{-1}$) & (mmag)  &  \\
		7 & 1.2309341(99) & 0.1967(52) & 0.356(4) & 35.44 \\ 
		4 & 0.703461(14) & 0.1364(52) & 0.792(6) & 13.84 \\ 
		8 & 1.406766(25) & 0.0794(52) & 0.247(10) & 15.75 \\ 
		9 & 1.582618(29) & 0.0669(52) & 0.249(12) & 15.13 \\ 
		11 & 1.934332(49) & 0.0401(52) & 0.910(21) & 12.17 \\ 
		6 & 1.055140(55) & 0.0356(52) & 0.609(23) & 5.37 \\ 
		10 & 1.758472(60) & 0.0324(52) & 0.178(25) & 8.76 \\ 
		12 & 2.110059(86) & 0.0227(52) & 0.259(36) & 7.90 \\ 
		3 & 0.527578(26) & 0.0755(52) & 0.531(30) & 4.57 \\ 
		\hline
	\end{tabular*}
\end{table}

\begin{figure}
	\includegraphics[width=\columnwidth]{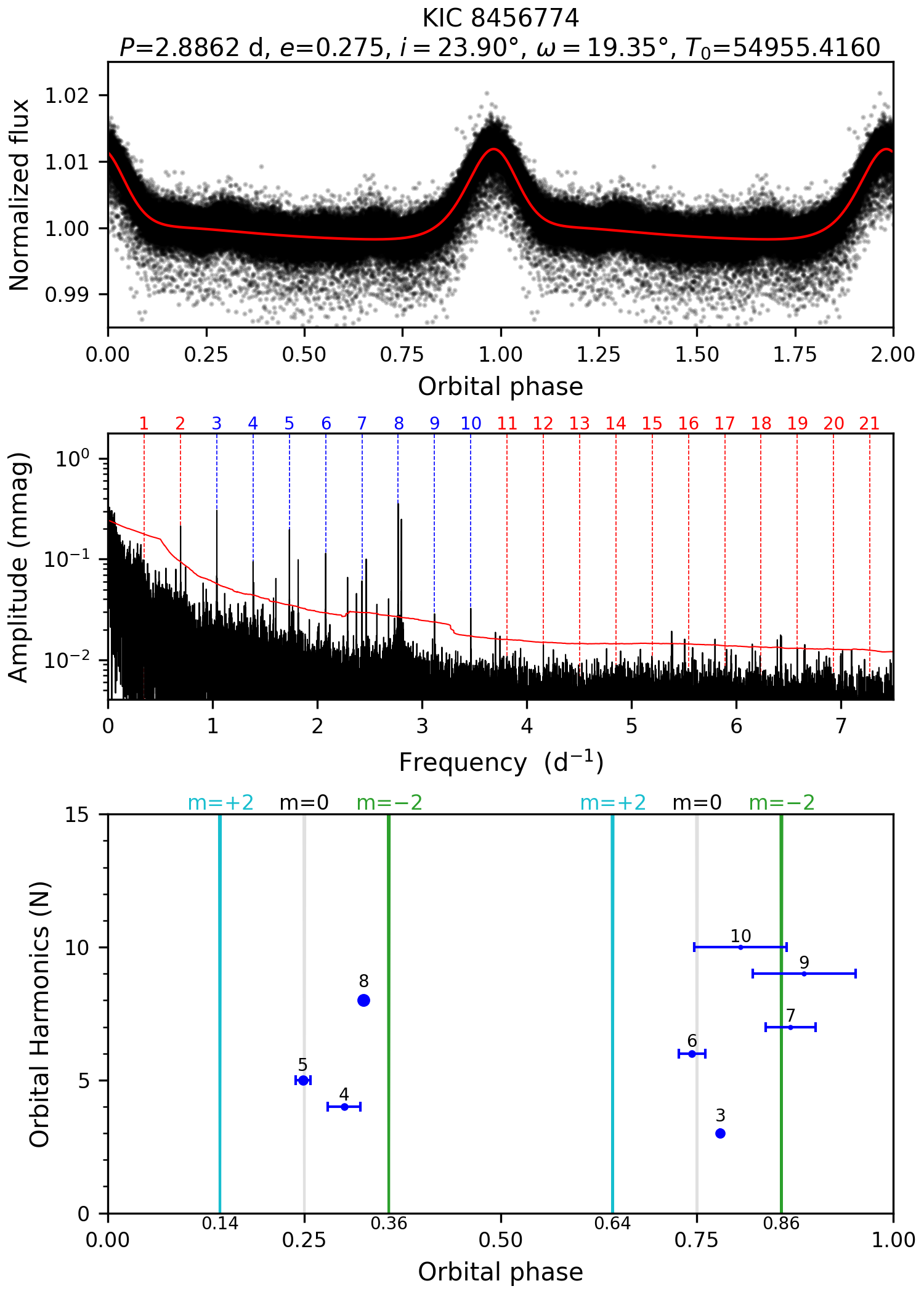}
	\caption{Same as Fig. \ref{fig:3547874} for KIC 8456774.
		\label{fig:8456774}}
\end{figure}

\subsubsection{KIC 8456774 (Figure \ref{fig:8456774}; Table \ref{tab:8456774})}
This system has a short-period of near 3 days, with a low eccentricity of 0.27 and a low inclination of 23$^{\circ}$.9. \citetalias{2024ApJ...962...44L} reported the $n$ = 8, 3, 5, and 6 harmonics are prominent TEOs shown in the middle panel of Figure \ref{fig:8456774}. The bottom panel shows that the 3, 4, 5, and 6 harmonics agree with the $m=0$ mode, i.e., they meet the expectation for such a low inclination \citep{2020ApJ...888...95G}. The 8, 7, and 9 harmonics are close to the $m=-2$ mode. However, the uncertainty of $n$ = 10 harmonic is too significant to match the modes and should be ignored.

\begin{table}
	\caption{Orbital and TEO parameters of KIC 8456774.}
	\label{tab:8456774}
	\begin{tabular*}{\columnwidth}{l@{\hspace*{9pt}}l@{\hspace*{9pt}}l@{\hspace*{8pt}}l@{\hspace*{7pt}}l}
		\hline
		Orbital Parameter &  & \\
		\hline
		$P$(d) & 2.88624552(21) & \\
		$e$ & 0.27480(10) & \\
		$i$($^\circ$) & 23.897(4) & \\
		$\omega$($^\circ$) & 19.347(19) & \\
		$T_{_{0}}$(BJD$-$2,400,000) & 54955.416037(79) & \\
		\hline
		Harmonic number & Frequency & Amplitude & Phase & S/N \\
		n & (d$^{-1}$) & (mmag)  &  \\
		8 & 2.771776(13) & 0.358(12) & 0.325(5) & 53.32 \\ 
		3 & 1.039410(15) & 0.306(12) & 0.780(6) & 21.62 \\ 
		5 & 1.732333(23) & 0.198(12) & 0.248(10) & 22.60 \\ 
		6 & 2.078805(40) & 0.115(12) & 0.744(17) & 15.69 \\ 
		4 & 1.385810(48) & 0.094(12) & 0.301(20) & 8.57 \\ 
		7 & 2.425258(75) & 0.061(12) & 0.869(32) & 8.23 \\ 
		10 & 3.46467(14) & 0.033(12) & 0.806(59) & 7.57 \\ 
		9 & 3.11817(16) & 0.029(12) & 0.886(66) & 4.97 \\ 
		\hline
	\end{tabular*}
\end{table}

\begin{figure}
	\includegraphics[width=\columnwidth]{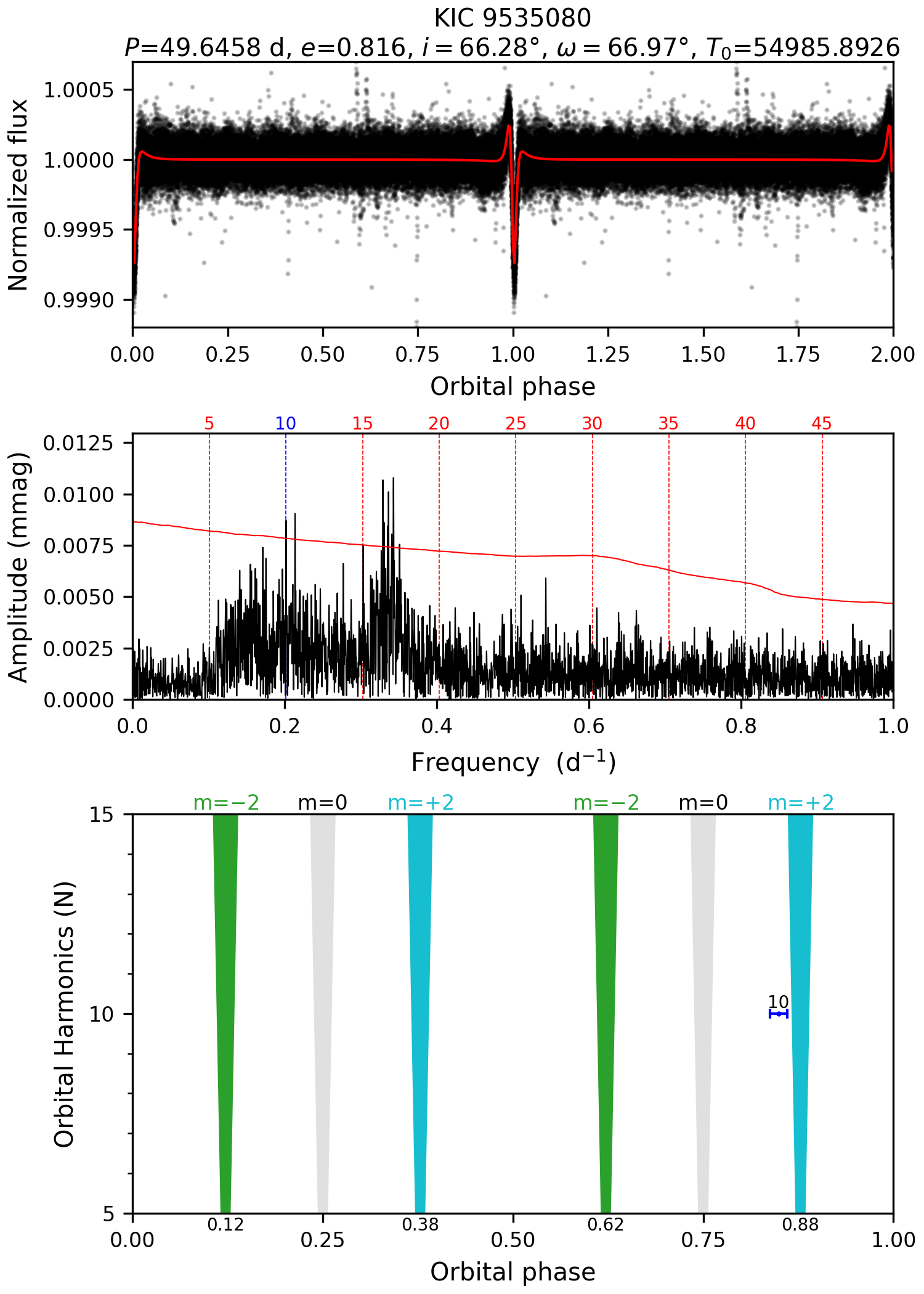}
	\caption{Same as Fig. \ref{fig:3547874} for KIC 9535080.
		\label{fig:9535080}}
\end{figure}

\subsubsection{KIC 9535080 (Figure \ref{fig:9535080}; Table \ref{tab:9535080})}
This is a HB system with a nearly 50-day period, with a high eccentricity of 0.816 and an inclination of 66$^{\circ}$ shown in Figure \ref{fig:9535080}. The $n$ = 10 harmonic is present in the middle panel. The bottom panel shows that its pulsation phase is close to the $m=2$ mode.

\begin{table}
	\caption{Orbital and TEO parameters of KIC 9535080.}
	\label{tab:9535080}
	\begin{tabular*}{\columnwidth}{l@{\hspace*{9pt}}l@{\hspace*{9pt}}l@{\hspace*{8pt}}l@{\hspace*{7pt}}l}
		\hline
		Orbital Parameter &  & \\
		\hline
		$P$(d) & 49.64577(13) & \\
		$e$ & 0.81643(50) & \\
		$i$($^\circ$) & 66.28(37) & \\
		$\omega$($^\circ$) & 66.97(36) & \\
		$T_{_{0}}$(BJD$-$2,400,000) & 54985.8926(35) & \\
		\hline
		Harmonic number & Frequency & Amplitude & Phase & S/N \\
		n & (d$^{-1}$) & (mmag)  &  \\
		10 & 0.201627(27) & 0.00879(63) & 0.849(11) & 4.45 \\ 
		\hline
	\end{tabular*}
\end{table}

\begin{figure}
	\includegraphics[width=\columnwidth]{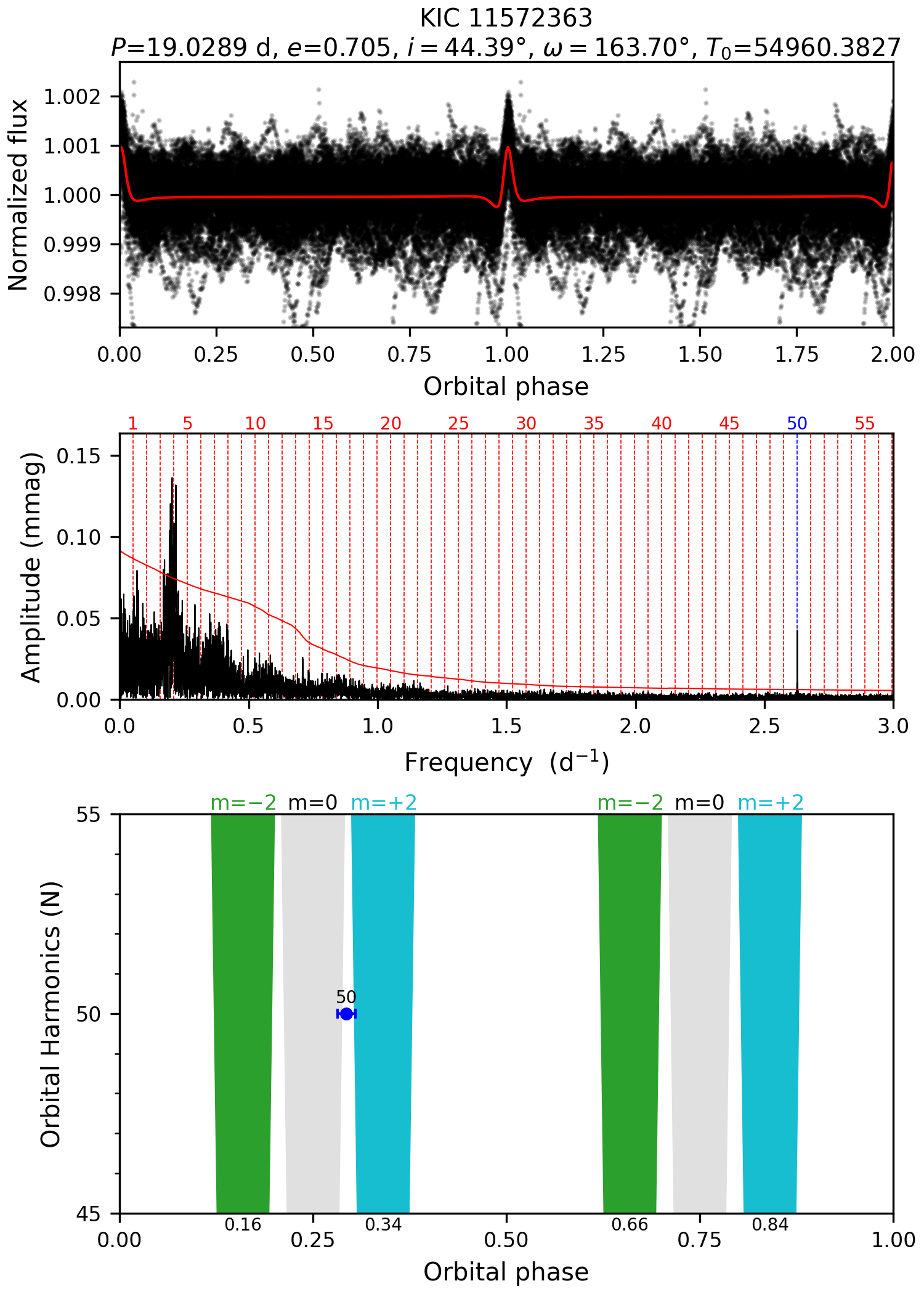}
	\caption{Same as Fig. \ref{fig:3547874} for KIC 11572363.
		\label{fig:11572363}}
\end{figure}

\subsubsection{KIC 11572363 (Figure \ref{fig:11572363}; Table \ref{tab:11572363})}
This is a nearly 20 day binary system, with an eccentricity of 0.7, an inclination of 44$^{\circ}$, and an argument of periastron $\omega$ of 164$^\circ$, as shown in Figure \ref{fig:11572363}. The $n$ = 50 harmonic is clearly visible in the middle panel. The bottom panel shows that its pulsation phase is close to the $m = 0$ or $m =-2$ modes. However, since the heartbeat signal has a small amplitude in the light curves in the top panel, the large uncertainty of $\omega$ is derived by the K95$^+$ model. Other modeling approaches may derive a more accurate mode in future work.

\begin{table}
	\caption{Orbital and TEO parameters of KIC 11572363.}
	\label{tab:11572363}
	\begin{tabular*}{\columnwidth}{l@{\hspace*{9pt}}l@{\hspace*{9pt}}l@{\hspace*{8pt}}l@{\hspace*{7pt}}l}
		\hline
		Orbital Parameter &  & \\
		\hline
		$P$(d) & 19.028898(25) & \\
		$e$ & 0.70500(56) & \\
		$i$($^\circ$) & 44.386(75) & \\
		$\omega$($^\circ$) & 163.70(26) & \\
		$T_{_{0}}$(BJD$-$2,400,000) & 54960.3827(19) & \\
		\hline
		Harmonic number & Frequency & Amplitude & Phase & S/N \\
		n & (d$^{-1}$) & (mmag)  &  \\
		50 & 2.627626(27) & 0.0425(31) & 0.293(12) & 28.32 \\ 
		\hline
	\end{tabular*}
\end{table}

\subsection{Systems with a misaligned spin orbit}\label{sec:misaligned}

\begin{figure}
	\includegraphics[width=\columnwidth]{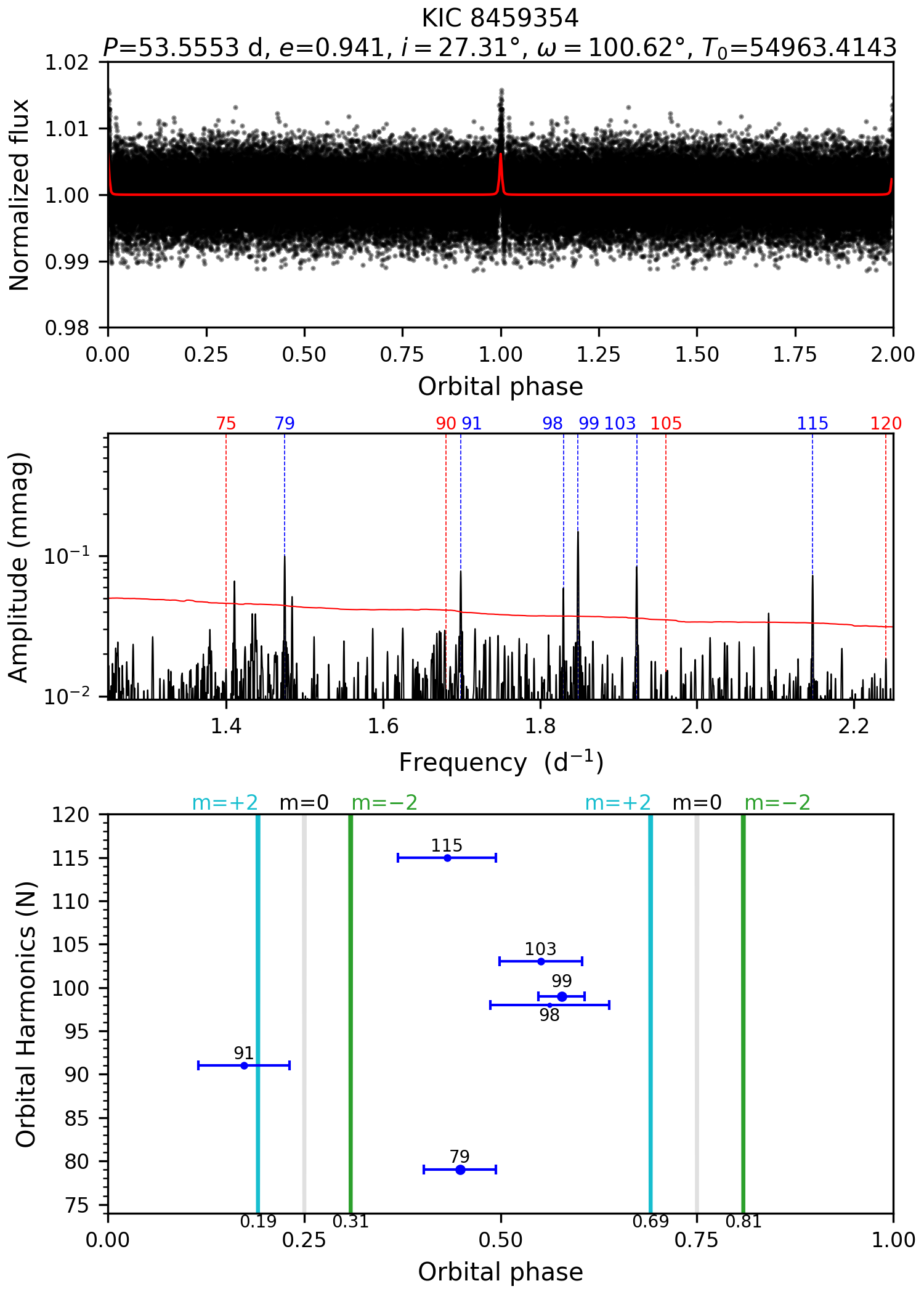}
	\caption{Same as Fig. \ref{fig:3547874} for KIC 8459354.
		\label{fig:8459354}}
\end{figure}

\subsubsection{KIC 8459354 (Figure \ref{fig:8459354}; Table \ref{tab:8459354})}
The orbital period is 53.56 days, the eccentricity $e$=0.94 and the inclination $i$=27$^{\circ}$.31. Such a low inclination will make the $m=0$ mode more visible according to \citet{2020ApJ...888...95G}. However, the bottom panel shows that its pulsation phases have large deviations ($>6\sigma$) from the adiabatic expectations. Given the highest eccentricity and long orbital period, we propose a strong spin-orbit misalignment in this HBS. Therefore, the TEOs in this system cannot be modeled with Eq. (\ref{equation:phi}). A further study of the $|m|=1$ modes for the misaligned cases \citep{2017MNRAS.472.1538F} is left for future work.

\begin{table}
	\caption{Orbital and TEO parameters of KIC 8459354.}
	\label{tab:8459354}
	\begin{tabular*}{\columnwidth}{l@{\hspace*{9pt}}l@{\hspace*{9pt}}l@{\hspace*{8pt}}l@{\hspace*{7pt}}l}
		\hline
		Orbital Parameter &  & \\
		\hline
		$P$(d) & 53.555278(22) & \\
		$e$ & 0.94122(4) & \\
		$i$($^\circ$) & 27.311(17) & \\
		$\omega$($^\circ$) & 100.624(44) & \\
		$T_{_{0}}$(BJD$-$2,400,000) & 54963.41426(20) & \\
		\hline
		Harmonic number & Frequency & Amplitude & Phase & S/N \\
		n & (d$^{-1}$) & (mmag)  &  \\
		99 & 1.84854(13) & 0.150(28) & 0.578(30) & 15.93 \\ 
		79 & 1.47503(20) & 0.097(28) & 0.448(46) & 8.74 \\ 
		103 & 1.92319(23) & 0.085(28) & 0.551(52) & 9.31 \\ 
		91 & 1.69919(26) & 0.077(28) & 0.173(58) & 7.75 \\ 
		115 & 2.14729(28) & 0.072(28) & 0.432(62) & 8.7 \\ 
		98 & 1.82986(34) & 0.059(28) & 0.562(76) & 6.08 \\ 
		\hline
	\end{tabular*}
\end{table}

\begin{figure}
	\includegraphics[width=\columnwidth]{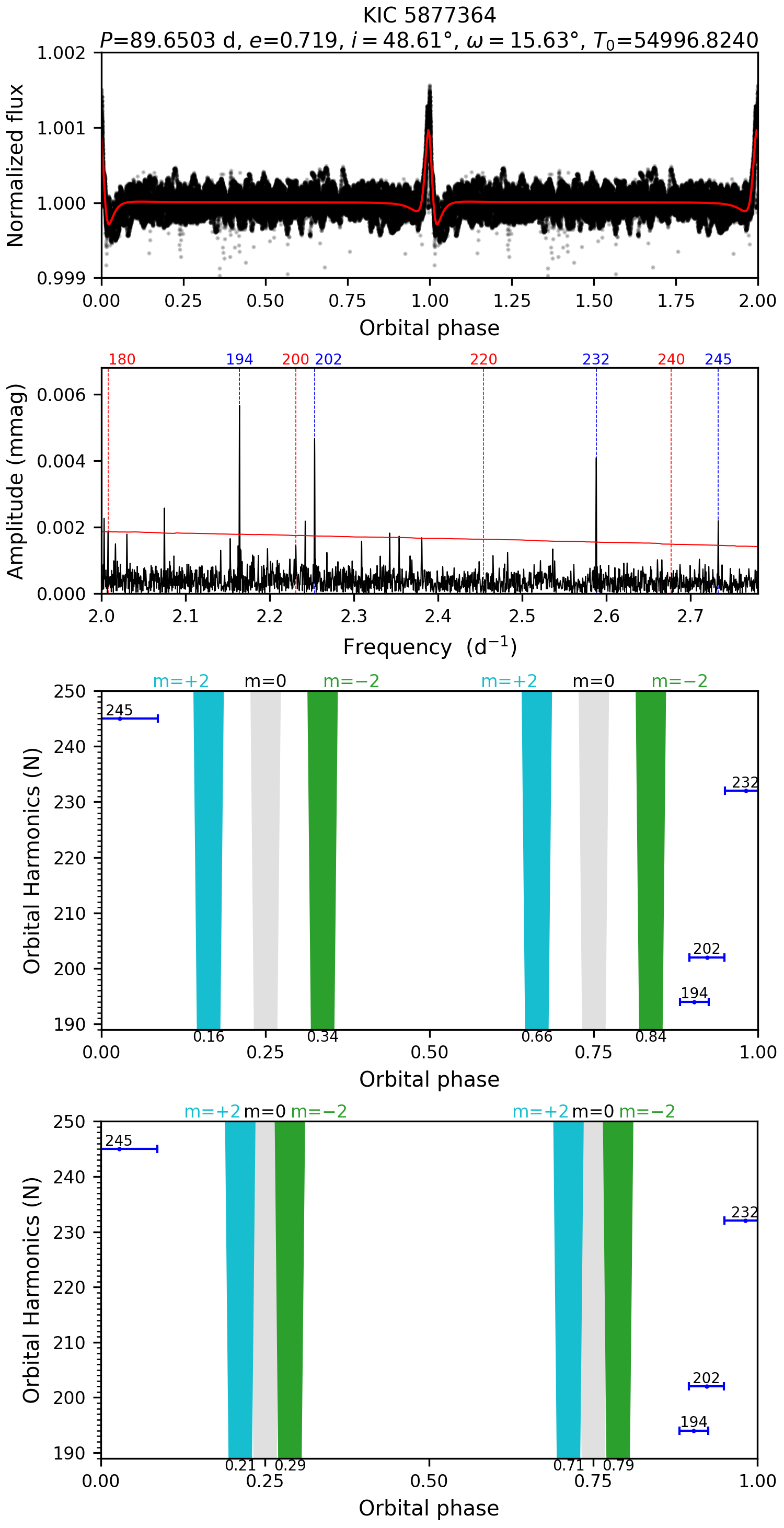}
	\caption{The first three panels are the same as in Fig. \ref{fig:3547874} for KIC 5877364. Bottom panel: the $m=\pm2$ strips correspond to the $\omega$ of 96$^\circ$.807 from \citet{2016ApJ...829...34S}.
		\label{fig:5877364}}
\end{figure}

\subsubsection{KIC 5877364 (Figure \ref{fig:5877364}; Table \ref{tab:5877364})}
\citet{2016ApJ...829...34S} reported a long orbital period of 89 days, a high eccentricity $e$=0.8875, and an argument of periastron $\omega$ of 96$^\circ$.807. We obtained a similar eccentricity $e$=0.719 and a different argument of periastron $\omega$=15$^\circ$.6 from \citetalias{2023ApJS..266...28L}. \citet{2022MNRAS.511..560S} reported the r-mode features and a rotation frequency of 0.30d$^{-1}$ in this system. In addition, \citetalias{2024ApJ...962...44L} reported four TEOs in this system. The $n$ = 194, 202, and 232 harmonics are prominent TEOs shown in the second panel of Figure \ref{fig:5877364}. The third panel shows that the $n$ = 202, 232 and 245 harmonics show large deviations ($>2\sigma$) from the adiabatic expectations according to the $\omega$ from \citetalias{2023ApJS..266...28L}. The $n$ = 194 harmonic is also have a larger deviation. We suggest that a spin-orbit misalignment also in this system, because of the higher eccentricity and long orbital period.

In addition, the bottom panel shows the $m=\pm2$ strips using the $\omega$ of 96$^\circ$.807 from \citet{2016ApJ...829...34S}. As can be seen, they are very close to the gray strips. However, the large deviations from the theoretical phases are more obvious, further supporting the expectation of a spin-orbit misalignment.

\begin{table}
	\caption{Orbital and TEO parameters of KIC 5877364.}
	\label{tab:5877364}
	\begin{tabular*}{\columnwidth}{l@{\hspace*{9pt}}l@{\hspace*{9pt}}l@{\hspace*{8pt}}l@{\hspace*{7pt}}l}
		\hline
		Orbital Parameter &  & \\
		\hline
		$P$(d) & 89.650258(77) & \\
		$e$ & 0.71930(9) & \\
		$i$($^\circ$) & 48.615(13) & \\
		$\omega$($^\circ$) & 15.632(31) & \\
		$T_{_{0}}$(BJD$-$2,400,000) & 54996.8240(10) & \\
		\hline
		Harmonic number & Frequency & Amplitude & Phase & S/N \\
		n & (d$^{-1}$) & (mmag)  &  \\
		194 & 2.163987(52) & 0.00583(81) & 0.903(22) & 12.70 \\ 
		202 & 2.253208(63) & 0.00484(81) & 0.922(27) & 10.76 \\ 
		232 & 2.587824(74) & 0.00409(81) & 0.981(31) & 10.56 \\ 
		245 & 2.73282(14) & 0.00222(81) & 0.028(58) & 6.04 \\ 
		\hline
	\end{tabular*}
\end{table}

\subsection{Systems that partially meet or close to expectations} \label{sec:part_meet}

\begin{figure}
	\includegraphics[width=\columnwidth]{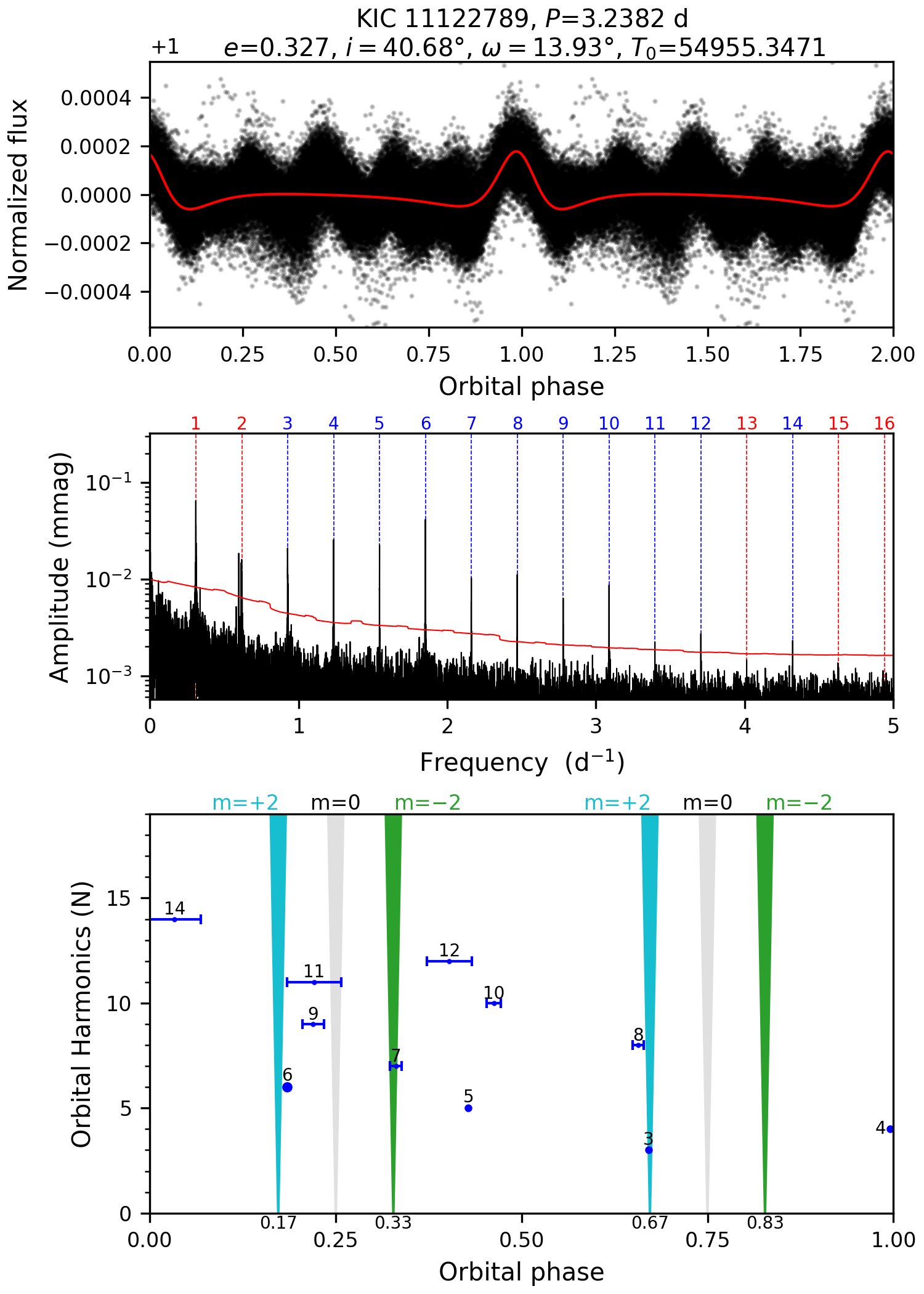}
	\caption{Same as Fig. \ref{fig:3547874} for KIC 11122789.
		\label{fig:11122789}}
\end{figure}

\subsubsection{KIC 11122789 (Figure \ref{fig:11122789}; Table \ref{tab:11122789})}
We obtained a short orbital period of 3 days, a low eccentricity $e$=0.327, an inclination of 41$^{\circ}$ and an argument of periastron $\omega$ of 14$^\circ$ in \citetalias{2023ApJS..266...28L}. The bottom panel of Figure \ref{fig:11122789} shows that the $n$ = 3, 6, and 8 harmonics are consistent with the $m=2$ mode; the $n$ = 7 and 12 harmonics are consistent with or close to the $m=-2$ mode; the $n$ = 9 and 11 harmonics are close to the $m=0$ mode. In addition, the $n$ = 4, 5, 10, and 14 harmonics show large deviations ($>5\sigma$) from the adiabatic expectations. However, the spin-orbit misalignment cannot explain these deviations, since other expected modes exist in the same system. Another explanation is that these harmonics cannot be the real TEOs, since there may be an imperfect fit in \citetalias{2023ApJS..266...28L}. Considering the top and middle panels, we cannot identify all the TEOs using only the Fourier transform analysis. This mode identification approach can be used to verify the TEOs identified by the Fourier spectrum. We suggest that if a harmonic does not fit the modes well, it may not be the TEO candidate and further investigation is needed.

\begin{table}
	\caption{Orbital and TEO parameters of KIC 11122789.}
	\label{tab:11122789}
	\begin{tabular*}{\columnwidth}{l@{\hspace*{9pt}}l@{\hspace*{9pt}}l@{\hspace*{8pt}}l@{\hspace*{7pt}}l}
		\hline
		Orbital Parameter &  & \\
		\hline
		$P$(d) & 3.2382374(22) & \\
		$e$ & 0.32664(47) & \\
		$i$($^\circ$) & 40.684(44) & \\
		$\omega$($^\circ$) & 13.93(19) & \\
		$T_{_{0}}$(BJD$-$2,400,000) & 54955.3471(12) & \\
		\hline
		Harmonic number & Frequency & Amplitude & Phase & S/N \\
		n & (d$^{-1}$) & (mmag)  &  \\
		6 & 1.8524487(63) & 0.03109(52) & 0.184(3) & 55.4 \\ 
		4 & 1.2350466(76) & 0.02583(52) & 0.996(3) & 28.71 \\ 
		5 & 1.5440340(86) & 0.02256(52) & 0.428(4) & 27.05 \\ 
		3 & 0.9255531(94) & 0.02083(52) & 0.670(4) & 18.88 \\ 
		8 & 2.470191(17) & 0.01125(52) & 0.657(7) & 19.81 \\ 
		7 & 2.161618(19) & 0.01043(52) & 0.331(8) & 15.01 \\ 
		10 & 3.087879(23) & 0.00866(52) & 0.463(10) & 17.87 \\ 
		9 & 2.779435(34) & 0.00635(52) & 0.220(14) & 12.19 \\ 
		12 & 3.705339(71) & 0.00276(52) & 0.403(30) & 6.18 \\ 
		14 & 4.322875(83) & 0.00234(52) & 0.033(35) & 5.6 \\ 
		11 & 3.396299(86) & 0.00227(52) & 0.221(36) & 4.81 \\ 
		\hline
	\end{tabular*}
\end{table}

\begin{figure}
	\includegraphics[width=\columnwidth]{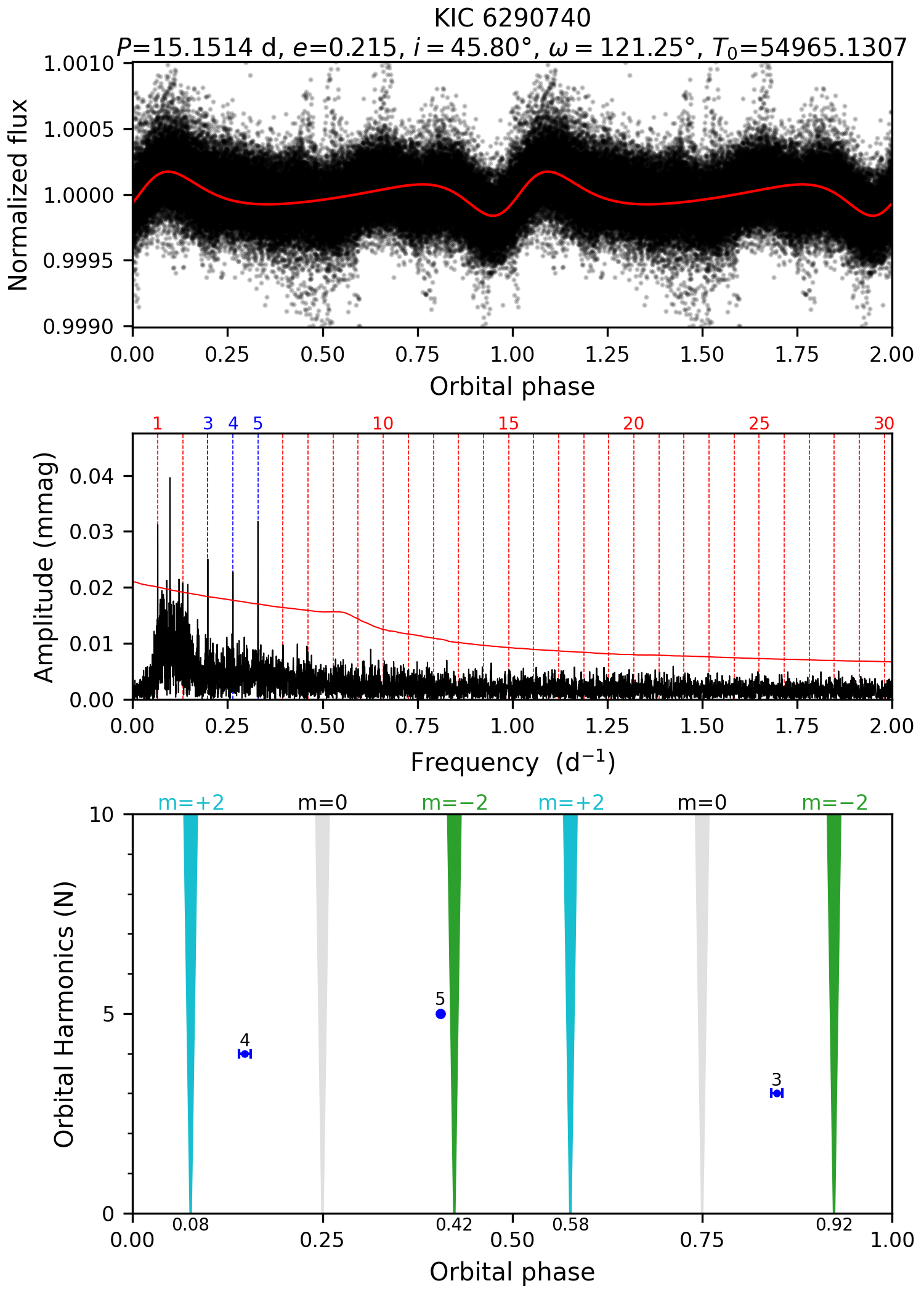}
	\caption{Same as Fig. \ref{fig:3547874} for KIC 6290740.
		\label{fig:6290740}}
\end{figure}

\subsubsection{KIC 6290740 (Figure \ref{fig:6290740}; Table \ref{tab:6290740})}
This is a 15 day HBS system, with a lower eccentricity of 0.21, an inclination of 45$^{\circ}$.8, and an argument of periastron $\omega$ of 121$^\circ$ shown in the top panel of Figure \ref{fig:6290740}. The bottom panel shows that the $n$=5 harmonic is in agrees with the $m=-2$ mode. However, the $n$ = 3 and 4 harmonics show large deviations ($>2\sigma$) from the adiabatic expectations. Given their lower amplitudes and $S/N$ in Table \ref{tab:6290740}, we suggest that they may be dominated by imperfect removal of the binary features. Therefore, they may not be considered as TEO candidates and are unsuitable for adiabatic expectations.

\begin{table}
	\caption{Orbital and TEO parameters of KIC 6290740.}
	\label{tab:6290740}
	\begin{tabular*}{\columnwidth}{l@{\hspace*{9pt}}l@{\hspace*{9pt}}l@{\hspace*{8pt}}l@{\hspace*{7pt}}l}
		\hline
		Orbital Parameter &  & \\
		\hline
		$P$(d) & 15.151353(99) & \\
		$e$ & 0.21454(99) & \\
		$i$($^\circ$) & 45.80(13) & \\
		$\omega$($^\circ$) & 121.25(29) & \\
		$T_{_{0}}$(BJD$-$2,400,000) & 54965.131(11) & \\
		\hline
		Harmonic number & Frequency & Amplitude & Phase & S/N \\
		n & (d$^{-1}$) & (mmag)  &  \\
		5 & 0.330129(13) & 0.0317(11) & 0.405(5) & 7.46 \\ 
		3 & 0.197976(17) & 0.0245(11) & 0.848(7) & 5.32\\ 
		4 & 0.264164(18) & 0.0232(11) & 0.148(7) & 5.16\\ 
		\hline
	\end{tabular*}
\end{table}

\begin{figure}
	\includegraphics[width=\columnwidth]{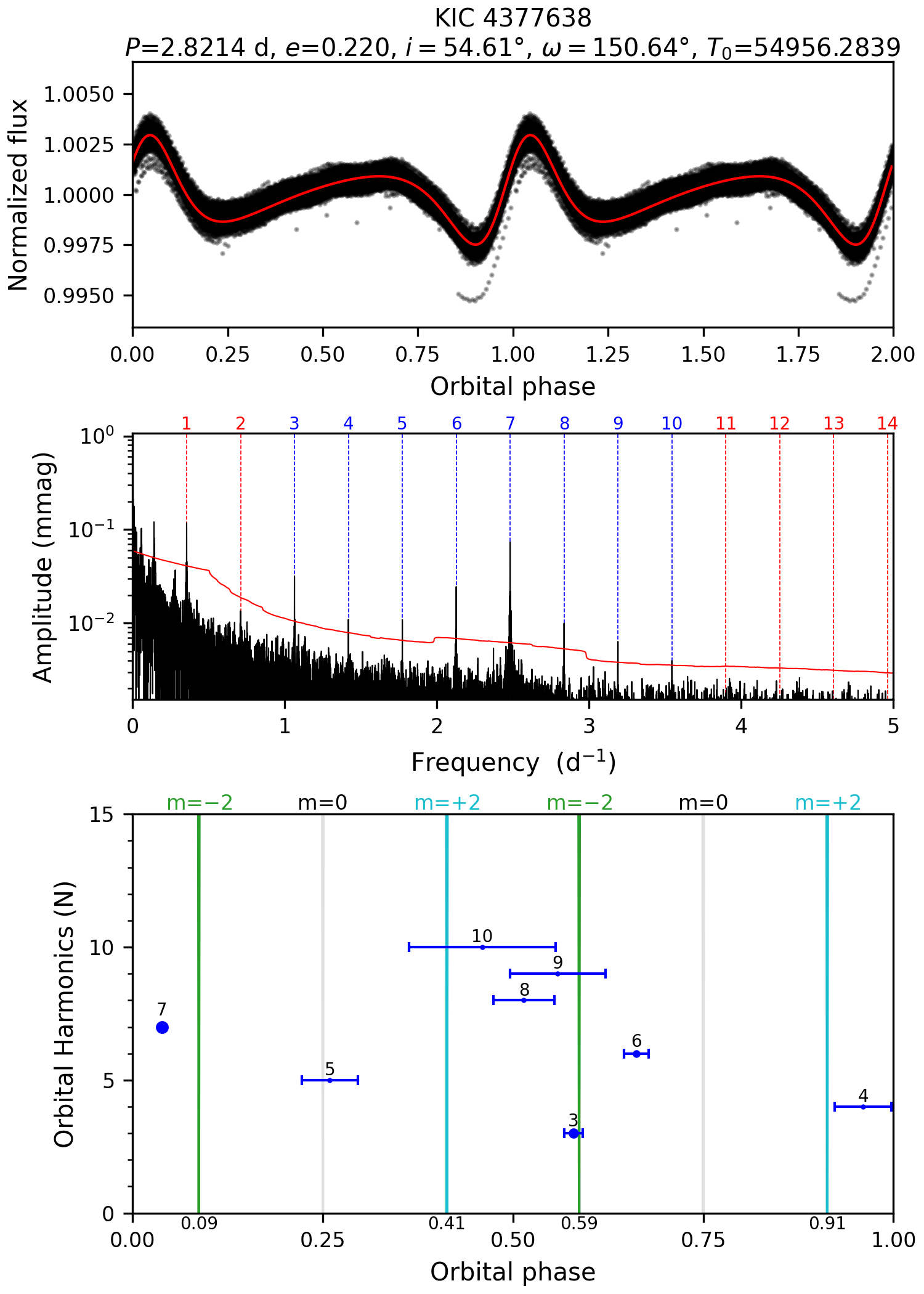}
	\caption{Same as Fig. \ref{fig:3547874} for KIC 4377638.
		\label{fig:4377638}}
\end{figure}

\subsubsection{KIC 4377638 (Figure \ref{fig:4377638}; Table \ref{tab:4377638})}
This is a short-period 2.8 day HBS system with a low eccentricity of 0.22 and an intermediate inclination of 54$^{\circ}$.6. Eight TEOs are reported in \citetalias{2024ApJ...962...44L}, and among those, $n$ = 7, 3, 6 are prominent. The bottom panel of Figure \ref{fig:4377638} shows that the phases of $n$ = 7, 3, 8, and 9 harmonics are consistent with or close to the $m=-2$ mode; the $n$ = 5 harmonics are consistent with the $m=0$ mode; the $n$ = 4 and 10 harmonics are close to the $m=2$ mode. However, the $n$ = 6 harmonic shows a large deviation ($>2\sigma$) from the adiabatic expectations. Given the large amplitude of the $n$ = 6 harmonic, we suggest that a travelling wave can explain it. The similar scenarios also exist in KOI-54 \citep{2014MNRAS.440.3036O}.

\begin{table}
	\caption{Orbital and TEO parameters of KIC 4377638.}
	\label{tab:4377638}
	\begin{tabular*}{\columnwidth}{l@{\hspace*{9pt}}l@{\hspace*{9pt}}l@{\hspace*{8pt}}l@{\hspace*{7pt}}l}
		\hline
		Orbital Parameter &  & \\
		\hline
		$P$(d) & 2.82141983(21) & \\
		$e$ & 0.22004(4) & \\
		$i$($^\circ$) & 54.609(11) & \\
		$\omega$($^\circ$) & 150.643(16) & \\
		$T_{_{0}}$(BJD$-$2,400,000) & 54956.28394(10) & \\
		\hline
		Harmonic number & Frequency & Amplitude & Phase & S/N \\
		n & (d$^{-1}$) & (mmag)  &  \\
		7 & 2.480888(16) & 0.0735(25) & 0.038(5) & 47.83 \\ 
		3 & 1.063304(37) & 0.0320(25) & 0.580(12) & 12.40 \\ 
		6 & 2.126581(49) & 0.0243(25) & 0.662(16) & 14.29 \\ 
		5 & 1.77220(11) & 0.0107(25) & 0.259(37) & 6.79 \\ 
		4 & 1.41789(11) & 0.0106(25) & 0.960(37) & 5.40 \\ 
		8 & 2.83540(12) & 0.0098(25) & 0.514(40) & 7.62 \\ 
		9 & 3.18978(19) & 0.0063(25) & 0.559(63) & 6.66 \\ 
		10 & 3.54428(29) & 0.0041(25) & 0.460(96) & 4.60 \\ 
		\hline
	\end{tabular*}
\end{table}

\begin{figure}
	\includegraphics[width=\columnwidth]{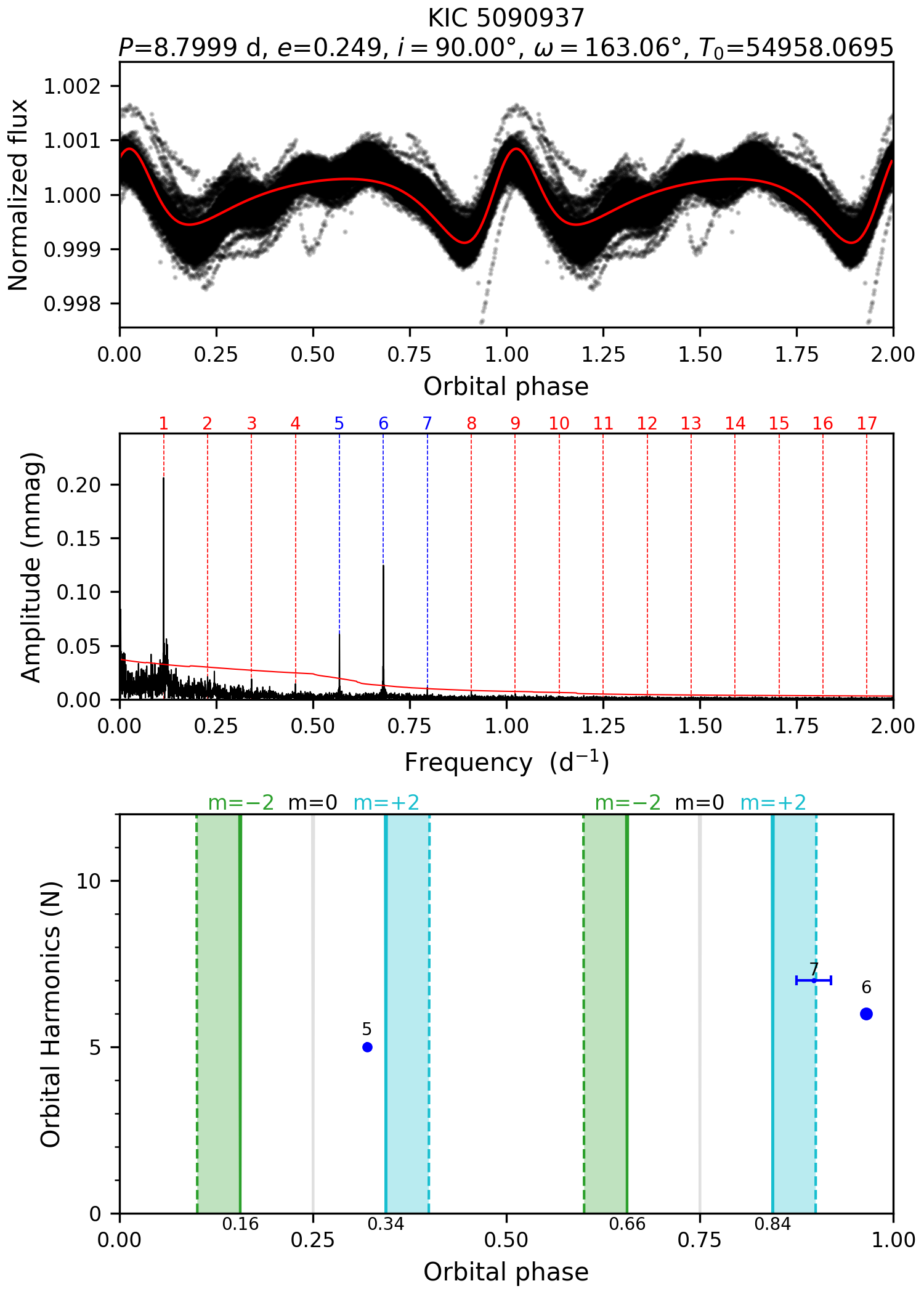}
	\caption{Same as Fig. \ref{fig:3547874} for KIC 5090937.
		\label{fig:5090937}}
\end{figure}

\subsubsection{KIC 5090937 (Figure \ref{fig:5090937}; Table \ref{tab:5090937})}
\citet{2016ApJ...829...34S} reported an orbital period of 8.8 days, an eccentricity $e$=0.241, and an argument of periastron $\omega$ of 153$^\circ$.071 according to radial velocities. We obtained eccentricity $e$=0.25, $\omega$=163$^\circ$.059 from \citetalias{2023ApJS..266...28L}. However, the inclination of this HBS may be overestimated (\citetalias{2023ApJS..266...28L}). \citet{2022MNRAS.511..560S} reported the Rossby mode (r-mode) features of this system. In addition, clear TEOs can also be seen in the top panel of Figure \ref{fig:5090937}. The $n$ = 6 and 5 harmonics are prominent TEOs, shown in the middle panel. In the bottom panel, the dashed lines represent the theoretical pulsation phases of $\omega$=153$^\circ$.071 from \citet{2016ApJ...829...34S}, and the lighter-colored strips represent the corresponding deviations. The $n$ = 7 harmonic is consistent with the $m=2$ mode. However, the $n$ = 5 and 6 harmonic show a large deviation ($>2\sigma$) from the adiabatic expectation (the dashed lines). Given their large amplitude, they are also expected to be travelling waves rather than standing waves. 

\begin{table}
	\caption{Orbital and TEO parameters of KIC 5090937.}
	\label{tab:5090937}
	\begin{tabular*}{\columnwidth}{l@{\hspace*{9pt}}l@{\hspace*{9pt}}l@{\hspace*{8pt}}l@{\hspace*{7pt}}l}
		\hline
		Orbital Parameter &  & \\
		\hline
		$P$(d) & 8.7998816(35) & \\
		$e$ & 0.24862(8) & \\
		$i$($^\circ$) & 90.000(63) & \\
		$\omega$($^\circ$) & 163.059(28) & \\
		$T_{_{0}}$(BJD$-$2,400,000) & 54958.06945(58) & \\
		\hline
		Harmonic number & Frequency & Amplitude & Phase & S/N \\
		n & (d$^{-1}$) & (mmag)  &  \\
		6 & 0.6818365(45) & 0.1240(15) & 0.965(2) & 38.79 \\ 
		5 & 0.5681886(95) & 0.0590(15) & 0.320(4) & 12.45 \\ 
		7 & 0.795427(53) & 0.0105(15) & 0.897(23) & 4.43 \\ 
		\hline
	\end{tabular*}
\end{table}

\begin{figure}
	\includegraphics[width=\columnwidth]{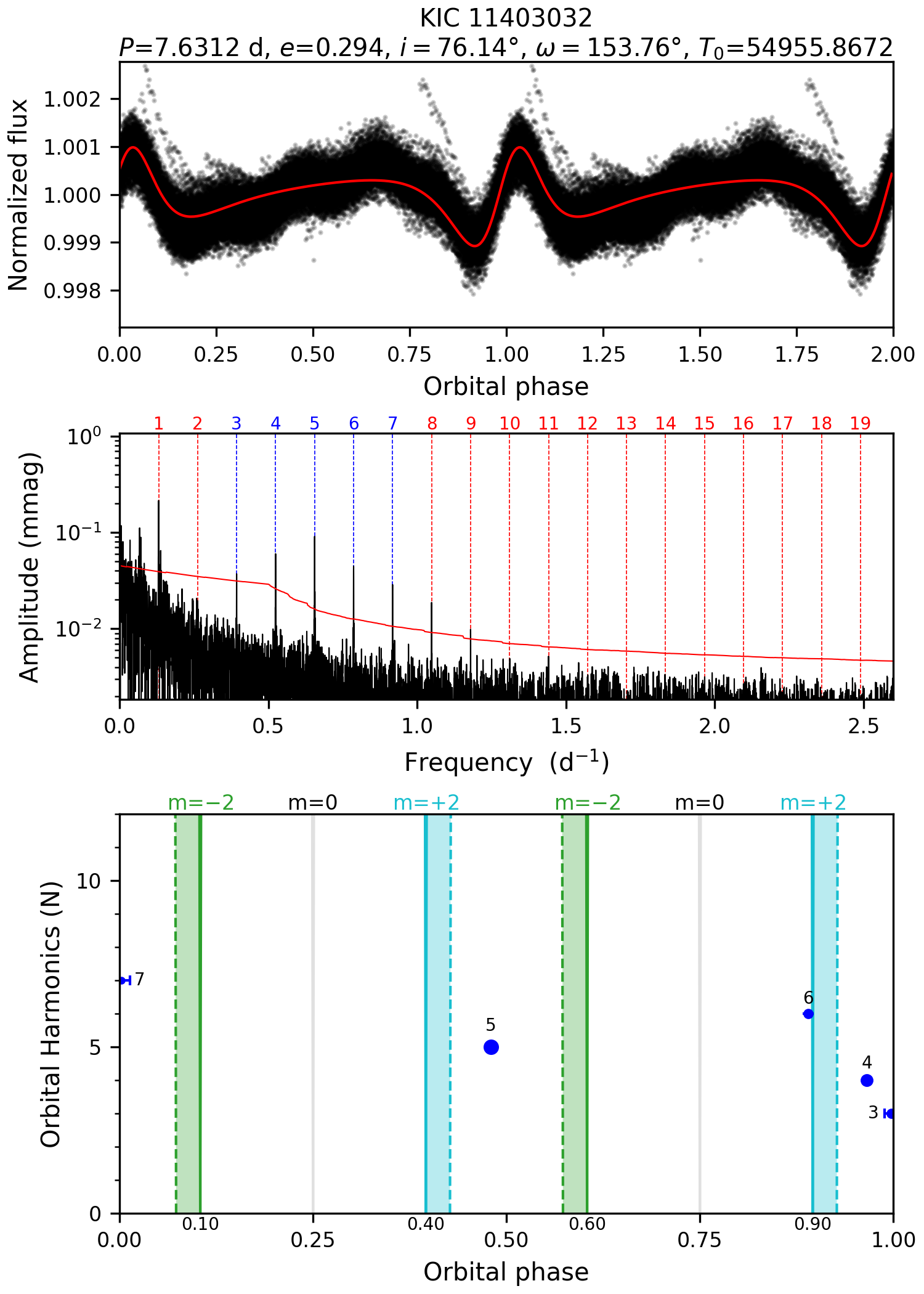}
	\caption{Same as Fig. \ref{fig:3547874} for KIC 11403032.
		\label{fig:11403032}}
\end{figure}

\subsubsection{KIC 11403032 (Figure \ref{fig:11403032}; Table \ref{tab:11403032})}
\citet{2016ApJ...829...34S} reported a short orbital period of 7.6 days, an eccentricity $e$=0.288, and an argument of periastron $\omega$ of 148$^\circ$.144. We obtained the eccentricity $e$=0.294, $i$=76$^\circ$.14, and $\omega$=153$^\circ$.757 from \citetalias{2023ApJS..266...28L}. \citet{2022MNRAS.511..560S} reported the r-mode features and the rotation period of 10.5 d in this system. In addition, clear TEOs can also be seen in the top panel of Figure \ref{fig:11403032}. The dashed lines in the bottom panel represent the theoretical pulsation phases of $\omega$=148$^\circ$.144, and the lighter-colored strips represent the corresponding deviations. It shows that the $n$ = 6, 5, and 4 harmonics are consistent with or close to the $m=2$ mode. However, the $n$ = 7 and 3 harmonics show large deviations ($>2\sigma$) from the adiabatic expectations. Because of their large amplitudes, they may be travelling waves rather than standing waves.

\begin{table}
	\caption{Orbital and TEO parameters of KIC 11403032.}
	\label{tab:11403032}
	\begin{tabular*}{\columnwidth}{l@{\hspace*{9pt}}l@{\hspace*{9pt}}l@{\hspace*{8pt}}l@{\hspace*{7pt}}l}
		\hline
		Orbital Parameter &  & \\
		\hline
		$P$(d) & 7.6312041(26) & \\
		$e$ & 0.29355(10) & \\
		$i$($^\circ$) & 76.144(86) & \\
		$\omega$($^\circ$) & 153.757(30) & \\
		$T_{_{0}}$(BJD$-$2,400,000) & 54955.86722(48) & \\
		\hline
		Harmonic number & Frequency & Amplitude & Phase & S/N \\
		n & (d$^{-1}$) & (mmag)  &  \\
		5 & 0.6551794(82) & 0.0907(20) & 0.480(3) & 22.66 \\ 
		4 & 0.524152(12) & 0.0597(20) & 0.966(5) & 9.23 \\ 
		6 & 0.786248(17) & 0.0443(20) & 0.890(7) & 14.13 \\ 
		7 & 0.917282(26) & 0.0290(20) & 0.002(11) & 10.79 \\ 
		3 & 0.393129(20) & 0.0379(20) & 0.997(8) & 4.73 \\ 
		\hline
	\end{tabular*}
\end{table}

\section{Discussions} \label{sec:discussion}
This work identifies the pulsation mode of TEOs using eq. (\ref{equation:phi}). The theoretical phase depends on the argument of periastron $\omega$. It is worth noting that there are differences in the $\omega$ obtained by the different approaches (see Table 3 in \citetalias{2023ApJS..266...28L}). The K95$^+$ model does not take into account other proximity effects, such as the irradiation/reflection effect and Doppler beaming \citep{2021A&A...647A..12K}. While other models that account for proximity effects may produce more reliable results. In addition, the same model may produce different results due to different detrending methods for the data. Therefore, a comprehensive consideration of these different values is required for a more reliable model identification.

In section \ref{sec:rst}, we study the phases and amplitudes of TEOs in fourteen HBSs. We find that the median deviations of the harmonic phases from the expected adiabatic phases are less than $2\sigma$ in most systems. However, in some systems there are some larger deviations from the expected adiabatic phases. \citet{2020ApJ...888...95G} have addressed two major reasons for the deviations. One is the non-adiabaticity of pulsations, and the other is spin-orbit misalignment. The largest deviations ($>6\sigma$) in KIC 8459354 can be explained by the major reason of spin-orbit misalignment. Since the high eccentricity ($e$=0.941) and the long orbital period ($P$=53.5553 d) of this system, the system is not yet tidally synchronized, nor are the spin and orbit axes aligned. Therefore, the large deviation in this system is not surprising. KIC 5877364 also has a similar scenario.

In addition, KIC 11122789 is an interesting sample in this work. Some modes have large deviations, while others agree well with the theoretical phases. However, the deviations cannot be explained by the non-adiabaticity of pulsations or the spin-orbit misalignment. Another explanation is that these harmonics may be excluded from the TEO candidates due to an imperfect fit. Given the top and middle panels in Figure \ref{fig:11122789}, the TEOs may not be identified by Fourier transform analysis alone. We suggest that the mode identification methodology can be a complementary approach to verify the TEOs identified by the Fourier spectrum. When the TEO candidates are identified by the Fourier analysis, further phase and mode identification is required. The harmonic can be considered a TEO candidate if the modes fit well. The large deviation harmonics in KIC 6290740 can also be explained by this reason. 

However, for KIC 4377638, the large deviation harmonics also have the large amplitudes; this cannot be explained by the imperfect removal. We suggest that these are travelling waves that do not satisfy the assumption in Section \ref{sec:analysis}. The large deviation harmonics in KIC 5090937 and KIC 11403032 can also be explained by this reason.

Furthermore, both the $m = 0$ and $m = 2$ modes can exist in the same system with a medium to high inclination. Such systems include KIC 4377638, KIC 5960989, KIC 8264510, and KIC 11122789.

Besides the two reasons addressed by \citet{2020ApJ...888...95G}, we propose another reason here: the apsidal motion. Note that KIC 3749404 is one of the large deviation systems in their paper. An interesting feature of this HBS is the rapid apsidal motion reported by \citet{2016MNRAS.463.1199H}. Since the apsidal motion is present, the argument of periastron $\omega$ varies with time. On the other hand, there is a positive correlation between $\omega$ and $T_{_{0}}$ in most of the HBSs (see Fig. 2 in \citetalias{2023ApJS..266...28L} and their online material). Then $T_{_{0}}$ also varies with time in the HBS with apsidal motion. However, the phases are sensitive to $T_{_{0}}$ in the Fourier spectrum analysis. Therefore, the derived phases may not be correct in an apsidal motion system. This could be a reason for the large deviations in KIC 3749404. For our samples, further investigation of the apsidal motion is needed in future work.

\section{Summary and Conclusions}\label{sec:conclusions}
Based on our previous work, \citetalias{2023ApJS..266...28L} and \citetalias{2024ApJ...962...44L}, we identify the pulsation phases and mode of TEOs in fourteen Kepler HBSs using the approach of \citet{2014MNRAS.440.3036O} and \citet{2020ApJ...888...95G}. Most pulsation phases of most systems can be explained by the dominant being $l=2$, $m=0$, or $\pm2$ spherical harmonic. Among these samples, KIC 8459354 shows the largest deviations ($>6\sigma$) from the adiabatic expectations. We suggest that this deviation can be explained by the spin-orbit misalignment due to its high eccentricity and long orbital period. KIC 5877364 also has a similar scenario. For KIC 11122789 and KIC 6290740, the TEO phases with large deviations may not be considered real TEOs. These harmonics may also not be considered TEO candidates. Furthermore, this approach can be used to inversely verify the TEO candidates derived by the Fourier analysis. For KIC 4377638, the large deviation harmonic with the largest amplitude is expected to be a travelling wave. Similar scenarios exist in two other systems: KIC 5090937 and KIC 11403032. In addition, besides the two reasons mentioned in \citet{2020ApJ...888...95G}, the apsidal motion could be another reason for the large deviations. For other spin-orbit misalignment systems, a further study of the $|m|=1$ modes approach \citep{2017MNRAS.472.1538F} is left for future work.

\section*{Acknowledgements}

This work is partly supported by the International Cooperation Projects of the National Key R$\&$D Program (No. 2022YFE0127300), the National Natural Science Foundation of China (Nos. 11933008 and 12103084), the Basic Research Project of Yunnan Province (Grant Nos. 202201AT070092 and 202301AT070352), the Science Foundation of Yunnan Province (No. 202401AS070046), and the Yunnan Revitalization Talent Support Program. We would like to thank \citet{2016AJ....151...68K} for their excellent work and for sharing the results of their work. The Kepler mission is funded by the NASA Science Mission Directorate. We are grateful to the Kepler teams for their support and hard work. We are grateful to the anonymous reviewer for constructive comments that have been a source of improvement for this manuscript.

\section*{Data Availability}
The data underlying this article are available in the \citet{2016AJ....151...68K} catalog, at \url{http://keplerEBs.villanova.edu}
 


\bibliographystyle{mnras}
\bibliography{example} 





\bsp	
\label{lastpage}
\end{document}